\documentclass[a4paper,fleqn,usenatbib]{mnras}
\usepackage{newtxtext,newtxmath}
\usepackage[T1]{fontenc}
\usepackage{ae,aecompl}
\usepackage{graphicx}   
\usepackage{amsmath}    
\usepackage{amssymb}    

\def\prl{PRL}
\def\prd{PRD}

\def\ssr{SSRv}
\def\lrl{LRL}
\def\pr{Phys. Rev.}

\title[Mass-redshift Degeneracy for the GW Sources in the Vicinity of SMBHs]{Mass-redshift Degeneracy for the Gravitational-wave Sources in the Vicinity of Supermassive Black Holes}
\author[X. Chen et al.]{
Xian Chen,$^{1,2}$\thanks{E-mail: xian.chen@pku.edu.cn}
Shuo Li$^{3}$
and Zhoujian Cao$^{4}$
\\
$^{1}$Astronomy Department, School of Physics, Peking University, 100871 Beijing, China\\
$^{2}$Kavli Institute for Astronomy and Astrophysics at Peking University, 100871 Beijing, China\\
$^{3}$National Astronomical Observatories and Key Laboratory of Computational Astrophysics, Chinese Academy of Sciences,\\
20A Datun Rd., Chaoyang District, Beijing 100012, China\\
$^{4}$Institute of Applied
Mathematics, Academy of Mathematics and Systems Science,
Chinese Academy of Sciences, Beijing 100190, China
}

\date{\today}
\pubyear{2018}

\begin{document}
\label{firstpage}
\pagerange{\pageref{firstpage}--\pageref{lastpage}}
\maketitle

\begin{abstract}
Retrieving the mass of a gravitational-wave (GW) source is a fundamental but
difficult problem because the mass is degenerate with redshift.  In astronomy,
three types of redshift exist, namely cosmological, Doppler, and gravitational
redshift, but the latter two are normally too weak to affect the observation.
In this Letter, we show that the current astrophysical models allow binary
black holes (BBHs) to merge within $10$ Schwarzschild radii of a supermassive
black hole (SMBH).  We find that in this case both the Doppler and
gravitational redshift are significant, and in the most extreme condition they
could increase the ``apparent'' black-hole mass and distance by a factor of
$1.9-3.4$. We show that such a factor is consistent with the distribution in
the distance-mass diagram of the ten BBHs detected so far by LIGO/Virgo. We
also discuss the difficulties of this redshift scenario caused by the low event
rate predicted by the current models, as well the potential solutions. 
\end{abstract}


\maketitle

\section{Introduction}

The Laser Interferometer Gravitational-wave Observatory (LIGO) and the Virgo
detectors have detected gravitational waves (GWs) from ten merging binary black
holes (BBHs) \citep{ligo18,ligo18bh}.  Interestingly, eight of these binaries
contain black holes (BHs) more massive than $20-30~M_\odot$. Such a large mass
has not been previously detected in  X-ray binaries
\citep{mcclintock14,corral16} and is $2-3$ times greater than the conventional
mass for stellar BHs, $10~M_\odot$.  Although the detected high mass can be
reconciled with the current BH formation models \citep{ligo16astro}, the
absence of $20-30~M_\odot$ BHs in X-ray binaries is more difficult to explain.

One possible solution is that we have significantly overestimated the masses of
the GW sources.  Its theoretical basis is the well-known degeneracy between
mass and redshift: By analysing GW data one derives only the redshifted mass
$m(1+z)$, which is greater than the rest mass $m$ by a redshift factor of $1+z$
\citep{schutz86}. Since cosmological redshift is omnipresent, it could have
caused the aforementioned discrepancy if what LIGO/Virgo have observed so far
were mostly strongly-lensed BBHs residing at large cosmological distances
\citep{smith18,broadhurst18}. This scenario, nevertheless, may not explain all
the eight detections because it predicts an inverse correlation between the
``apparent'' masses and distances of the BBHs which is contradictory to what
have been observed \citep{ligo18}. 

Besides cosmological redshift, in astronomy there exist also Doppler and
gravitational redshifts.  The latter two have not been considered as the cause
of the observed high masses because in the conventional view of BBH formation
the centre-of-mass (COM) velocity is small relative to the speed of light ($c$)
and the potential energy induced by the environment is negligible
\citep{ligo16astro,ama16}. However, the conventional view may not be complete
in light of the recent progress made by the studies of the stellar dynamics in
galactic nuclei.

Recent studies showed that the merger rate of BBHs is enhanced in nuclear star
clusters (NSCs) due to the presence of supermassive black holes (SMBHs).  The
enhancement is caused by the following factors.  (1) Stellar-mass BHs are more
easily retained in NSCs with SMBHs because the large escape velocity is large
\citep{miller09VL}.  (2) A dynamical effect called ``mass segregation''
increases the density of BHs around SMBHs
\citep{BW76,morris93,miralda00,FAK06a,hopman06ms,AlexanderHopman09,ASEtAl04}.
(3) SMBHs can tidally capture BBHs to bound orbits \citep{addison15,chen18}.
(4) Tidal perturbation by the SMBHs can enhance the merger rate of BBHs through
a mechanism called the ``Lidov-Kozai effect''
\citep{antonini12,prodan15,stephan16,VL16,liu17,petrovich17,bradnick17,hoang17,arca-sedda18}.
(5) If the environment is gas-rich, as will be the case in an active galactic
nucleus (AGN), BBHs can grow and merge more rapidly
\citep{syer91,bellovary16,bartos2017,stone2017,mckernan18}.

The results of these earlier studies suggest that it is possible a fraction of,
if not all, BBH mergers happen in the vicinity of SMBHs. This possibility
motivates us to revisit the problem of mass-redshift degeneracy, taking
especially the Doppler and gravitational redshift into account. Thoughout the
Letter, we adopt the convention $G=c=1$.

\section{Possible formation scenarios}\label{sec:location}

We first investigat the distance $r$ relative to a SMBH where BBH merger could
possibly happen.  We find in the literature two mechanisms that could deliver
BBHs to $r\lesssim10\,R_S$, where $R_S$ is the Schwarzschild radius of the
SMBH.  

The first mechanism is tidal capture \citep{addison15,chen18}.  The necessary
condition is that a BBH, initially far from and gravitationally unbound to a
SMBH, approaches the SMBH until it reashes a distance comparable to the ``tidal
radius'', $r_t:=a(M_3/m_{12})^{1/3}$, where $a$ is the semimajor axis of the
binary, $M_3$ is the mass of the SMBH and $m_{12}$ is the total mass of the
binary.  Moreover, the binary initially should have a small semimajor axis so
that after being captured it remains tightly bound to and not easily breakable
from the SMBH even though it repeatedly interacts with the background stars.
This condition for stability imposes to $a$ an upper limit $a_{\rm cri}$ which
depends on the relaxation timescale $T_{\rm rlx}$ of the NSC.

Using the $a_{\rm cri}$ derived in \citet{chen18} and assuming that the initial
pericentre of the COM of the BBH is $\xi r_t$ where $\xi$ is a factor of order
unity, we can derive the minimum distance of the captured BBHs as
\begin{equation}
\frac{r}{R_S}\simeq\frac{15(\xi q)^{3/8}}{(1+q)^{1/4}}
\left(\frac{T_{\rm rlx}}{10^9{\rm yrs}}\right)^{1/4}
\left(\frac{m_1}{10\,M_\odot}\right)^{1/2}
\left(\frac{M_3}{10^6\,M_\odot}\right)^{-3/4},\label{eq:Rp}
\end{equation}
where $m_1$ and $m_2$ are the masses of the two small BHs (we assume $m_1\ge
m_2$), and $q:=m_2/m_1$ is their mass ratio.  Only $1\%$ of the captured
binaries may coalesce at this distance according to preliminary results from
numerical simulations, and hence the event rate is too low to explain the
LIGO/Virgo detections unless the capture rate of BBHs has been significantly
underestimated in the current models \citep{chen18}.

The second mechanism is based on the scenario of BBH formation in AGN accretion
discs \citep{syer91,mckernan18}. After BBHs form in the disc, the
hydrodynamical interaction with the surrounding gas drives them to coalescence
\citep{stone2017,bartos2017}.  Several locations in the disc can trap stellar
BHs for a long time \citep{chakrabarti93,bellovary16}, and hence they are the
preferential places for BBH mergers.

In particular, when the accretion rate is comparable or exceeding the Eddington
limit, the inner part of the disc becomes super-Keplerian \citep{abramowicz13}
so that a stellar BH embedded in it gains angular momentum as it accretes from
the surrounding gas. This accretion compensates the angular-momentum loss via
the GW radiation generated by the orbit of the small BH around the SMBH
\citep{chakrabarti93}. An equilibrium could be achieved at a radius of
\begin{equation}
\frac{r}{R_S}\simeq15\,\dot{m}^{-2/7}\left(\frac{M_3}{10^9\,M_\odot}\right)^{-4/7}
\left(\frac{m}{20\,M_\odot}\right)^{4/7}\label{eq:r}
\end{equation}
\citep{chakrabarti93}, where $m$ is the mass of the small body embedded in the
disc and in our problem we set $m=m_{12}$. The accretion rate of the small
body, $\dot{m}$, is normalized by the Eddington rate. We note that the event
rate of BBH mergers in this scenario is unknown and deserves a future
investigation.
 
\begin{figure} 
\begin{center} 
\includegraphics[width=0.48\textwidth]{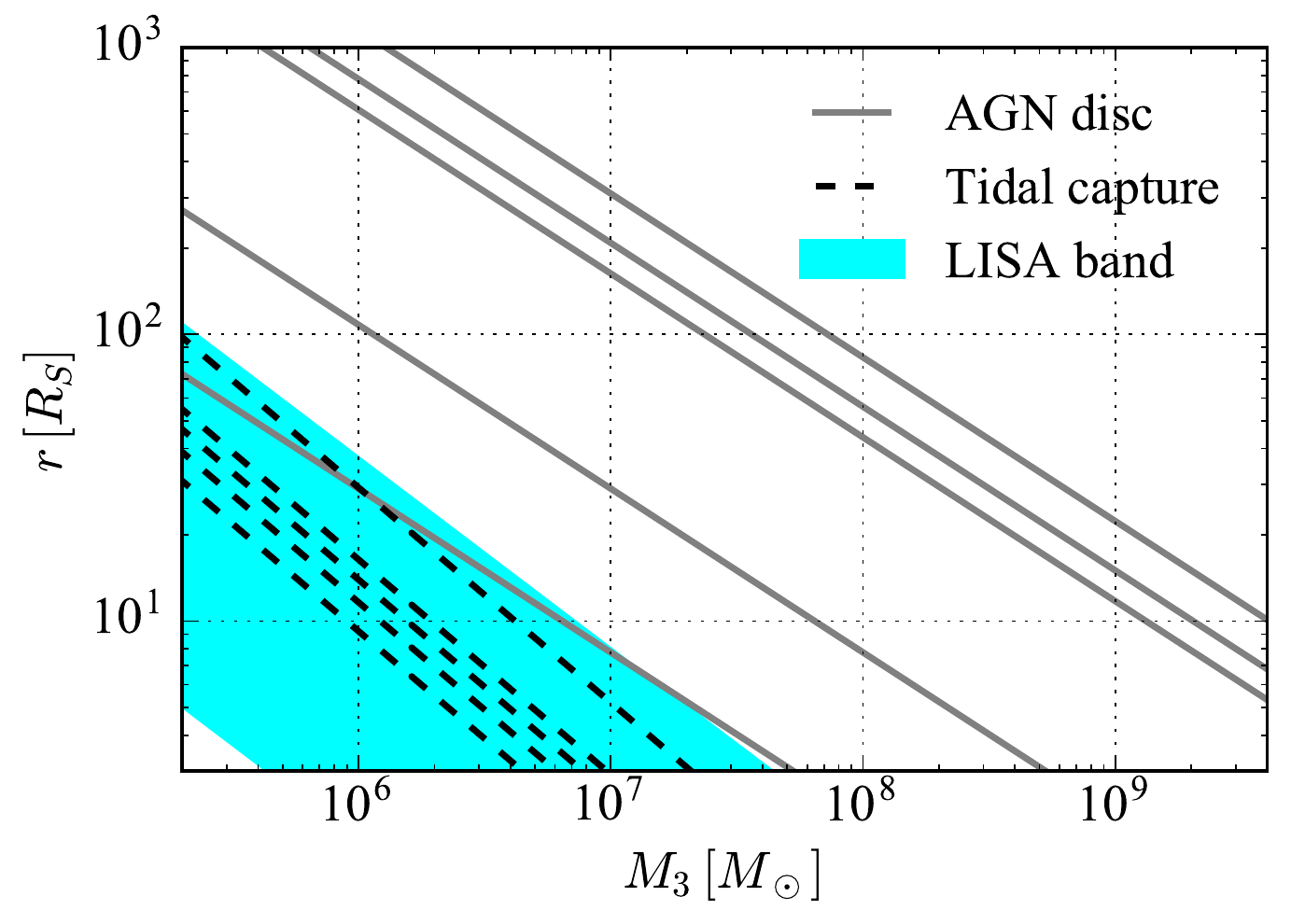}
\caption{Predicted location for BBH mergers as a function of the mass of the
central SMBH.  The fiducial parameters are $m_1=10M_\odot$, $q=1$, $T_{\rm
rlx}=10^9\,{\rm years}$, $\xi=2$ and $\dot{m}=1$.  The dashed lines are derived
from Equation~(\ref{eq:Rp}), which corresponds to the tidal-capture model
\citep{chen18}.  From top to bottom they refer to the results for,
respectively, (1) $m_1=20M_\odot$, (2) the fiducial model, (3) $\xi=1.3$, (4)
$q=0.3$ and (5) $T_{\rm rlx}=10^8\,{\rm years}$. The solid lines are computed
using Equation~(\ref{eq:r}), i.e., the equilibrium radius for a BBH embedded in
an AGN disc \citep{chakrabarti93}.  From top to bottom they correspond
to (1) $m_1=20M_\odot$, (2) the fiducial model, (3) $q=0.3$, (4) $\dot{m}=10^3$
and (5) $\dot{m}=10^5$. BBHs falling in the cyan-shaded area emit GWs in the
frequency band of $10^{-2}-10^{-4}$ Hz \citep[see, e.g.,][]{chen17b,han18} and
hence are detectable by a LISA-type interferometer.} \label{fig:Rp}
\end{center} 
\end{figure}

Figure~\ref{fig:Rp} illustrates the results from Equations~(\ref{eq:Rp}) and
(\ref{eq:r}).  We can see that for the tidal-capture model (dashed lines) to
produce BBHs at $r<10R_S$, the SMBHs should be about $10^{6-7}M_\odot$. We
notice that most of the SMBHs detected in the local universe fall in this mass
range \citep{kormendy13}. Regarding the accretion-disc model (solid lines), if
we assume $m_1\simeq m_2\simeq10\,M_\odot$, we find that $r\lesssim10\,R_S$
when $M_3\gtrsim2.0\times10^9\dot{m}^{-1/2}\,M_\odot$.  If we further consider
the results from the recent studies which showed that
$1\lesssim\dot{m}\lesssim10^5$ for the stellar BHs inside AGN accretion discs
\citep{inayoshi16,stone2017}, we can conclude that $M_3$ has to be at least
$6.3\times10^6\,M_\odot$ to produce BBH merge at $r\la10\,R_S$.

Figure~\ref{fig:Rp} also shows that both the tidally captured BBHs and those
most-rapidly accreting binaries in AGN discs are settled on such tightly bound
orbits around SMBHs that their orbital motion will generate GWs detectable by
the Laser Interferometer Space Antenne (LISA).  This result implies that future
joint observations of GWs using both LIGO/Virgo and LISA and help us identify
BBH mergers happening very close to SMBHs \citep{han18}.

\section{Effects of redshift on the measurement of mass}\label{sec:mass}

If BBHs coalesce within $10\,R_S$ of SMBHs, the Doppler and gravitational
redshift could significantly affect the interpretation of the observational
data.  In GW astronomy, the observables are the amplitude $h$ (in two
polarizations), frequency $f$, and the rate at which the signal chirps
$\dot{f}$. If space-time is flat without redshift effect, one can measure $f$
and $\dot{f}$ in the ``inspiral'' part of the waveform, where the BHs are far
apart so that they can be approximated by point masses, and infer from them an
intrinsic mass scale for the system,
\begin{equation}
{\cal M}:=\left(\frac{5f^{-11/3}\dot{f}}{96\pi^{8/3}}\right)^{3/5}.
\end{equation}
This is known as the ``chirp mass'' because it determines how the frequency
increases with time.  Theoretically, it depends on the masses of the two BHs
as ${\cal M}=(m_1m_2)^{3/5}(m_1+m_2)^{-1/5}$.  Together with the third
observable $h$, one can further infer the (luminosity) distance of the source
\citep{holz05}, using
\begin{equation}
d=(4{\cal M}/h)\left(\pi f{\cal M}\right)^{2/3}\label{eq:d}.
\end{equation}

Redshift, by definition, reduces the frequency of a wave from a value of $f$,
measured close to the source, to a value of $f_{o}=f(1+z)^{-1}$, where the
subscript $o$ indicates a quantity measured in the rest frame of an observer.
Correspondingly, the chirp rate changes from $\dot{f}$ to
$\dot{f}_{o}=\dot{f}(1+z)^{-2}$, where the additional $(1+z)$ factor comes
from time dilation. As a result, the only chirp mass that an observer can
construct from the observed waveform is
\begin{equation}
{\cal M}_{o}:=\left(\frac{5f_{o}^{-11/3}\dot{f}_{o}}{96\pi^{8/3}}\right)^{3/5}={\cal M}(1+z).
\end{equation}
It is greater than the intrinsic one by a redshift factor. Without an
independent measurement of $z$, there is no way of disentangling ${\cal M}$ and
$1+z$. This is a famous problem called ``the mass-redshift degeneracy'', and
this effect applies to all three kinds of redshift, namely, the cosmological,
Doppler and gravitational redshift.

\section{Effects on the measure of distance}\label{sec:distance}

Starting from the observables, the only possible way of constructing a distance
scale is via 
\begin{equation}
d_o:=(4{\cal M}_o/h_o)\left(\pi f_o{\cal M}_o\right)^{2/3},\label{eq:d_o}
\end{equation}
where $h_o$ is the observed amplitude. The distances of the LIGO/Virgo BBHs are
derived effectively in this way. Therefore, it is crucial, for the purpose of
this work, to understand what $d_o$ really is.

(1) Cosmological redshift: The expansion of the universe causes a redshift,
$z_{\rm cos}$, which increases with the transverse comoving distance
$d_C$.  The GW amplitude damps linearly with $d_C$, i.e.
\begin{equation} 
h_o=(4{\cal M}/d_C)\left(\pi f{\cal
M}\right)^{2/3}.\label{eq:h_cos} 
\end{equation}

Using the last equation to replace $h_o$ in Equation~(\ref{eq:d_o}) and
noticing that (a) $d_C(1+z_{\rm cos})$ is the luminosity distance $d_L$ in a
flat universe and (b) $f{\cal M}=f_o{\cal M}_o$, we find that $d_o=d_C(1+z_{\rm
cos})=d_L$.  Therefore, we recover the classical notion that GWs encode the
luminosity distance of the source \citep{schutz86}.

(2) Doppler redshift: The relative motion of a GW source with respect to an
observer also causes a shift to the observed frequency. It will be a redshift
if the source is receding relative to the observer and a blueshift if the
source is approaching.  For simplicity we restrict ourselves to the case in
which the COM of the BBH is moving at a constant velocity of $v$ away from the
observer. According to the theory of special relativity, the Doppler redshift is 
\begin{equation}
1+z_{\rm dop}=\gamma(1+\beta),\label{eq:z_dop}
\end{equation}
where $\beta=v/c$ and $\gamma=(1-\beta^2)^{-1/2}$ is the Lorentz factor. 

Now the total redshift becomes $1+z_{\rm tot}=(1+z_{\rm cos})(1+z_{\rm dop})$,
and, correspondingly, we have $f_{o}=f(1+z_{\rm tot})^{-1}$ and
$\dot{f}_{o}=\dot{f}(1+z_{\rm tot})^{-2}$. The redshifted mass, as can be
measured from GWs, consequently, becomes ${\cal M}_o={\cal M}(1+z_{\rm
cos})(1+z_{\rm dop})$.

As for $h_o$, numerical simulations show that to linear order it is not affect
by the Doppler effect \citep{gerosa16}.  We note that this result is derived
for plane waves and may not be applicable to spherical waves.  Therefore, in
the plane-wave approximation $h_o$ equals that $h$ in
Equation~(\ref{eq:h_cos}).  Using these new relations to replace ${\cal M}_o$,
$f_o$, and $h_o$ in Equation~(\ref{eq:d_o}), we find that $d_o=d_C(1+z_{\rm
cos})(1+z_{\rm dop})=d_L(1+z_{\rm dop})$.  This result shows that the apparent
distance $d_o$ is an overestimation of the real (luminosity) distance by a
factor of $1+z_{\rm dop}$. 

(3) Gravitational redshift: Waves originating from a deep gravitational
potential also get redshifted.  Since we are interested in GWs from the
vicinity of a SMBH, our gravitational potential reduces to that of a point
mass. The space-time surrounding it can be described by the
Schwarzschild metric assuming the simple case that the hole is not rotating.
In the following we restrict our discussions to $r>4M_3$, because it is a
limit imposed by the innermost bound orbit.

To derive the redshift for GWs, we first notice that $M_3/m_{12}>10^5$ for our
problem, since $M_3\gtrsim10^6\,M_\odot$ according to our earlier analysis and
we are interested in BBHs with $m_1\sim m_2\sim10\,M_\odot$.  This
disproportion has two consequences that significantly simplify our problem.
First, we are interested in the GWs emitted during the last few cycles of a
merger because BBHs enter the LIGO/Virgo band when their semimajor axes become
$a\lesssim 10m_{12}$. Therefore, we are in a regime where the curvature radius
of the background space-time, $\rho\sim\sqrt{r^3/M_3}>8M_3$ \citep{isaacson68},
is more than $10^5$ times greater than the wavelength of the gravitational
radiation, which is comparable to $a$. Second, the waveform of the merger
stretches over a time interval of $\Delta t\sim2\pi\sqrt{a^3/m_{12}}$, which
corresponds to a length of $\Delta l=c\Delta t$. This length is only a fraction
of $\Delta l/(2M_3)\lesssim\pi\sqrt{10^3}(m_{12}/M_3)\lesssim2\times10^{-3}$ of
the length scale of the Schwarzschild metric ($2M_3$). 

From the above arguments we conclude that we are in a short-wave limit. In this
case, GWs propagate in approximately the same way as light waves
\citep{isaacson68}. So we can use the redshift for photons, 
\begin{equation} 
1+z_{\rm gra}=(1-R_S/r)^{-1/2},\label{eq:z_gra} 
\end{equation}
to calculate the GW frequencies in the rest frame of the observer, who is
supposedly at infinity (``final observer'' hereafter). The last equation is
appropriate only for the case in which the source and observer are on the same
side of the SMBH.  Otherwise, the GWs have to circle around the SMBH to get to
the observer, in which case the waves could have reached a closer distance to
the SMBH.  Strong lensing effect is expected in the latter case
\citep{kocsis13}, but our analytical scheme cannot treat it properly.
Therefore, we take the simple case for illustrative purposes and find
\begin{equation}
1+z_{\rm tot}=(1+z_{\rm cos})(1+z_{\rm dop})(1+z_{\rm gra})
\end{equation}
and 

\begin{equation}
{\cal M}_o={\cal M}(1+z_{\rm cos})(1+z_{\rm dop})(1+z_{\rm gra}).
\end{equation}
  
As for the GW amplitude, we start our analysis from the point of view of an
intermediate observer at a small radial offset of $\Delta r\sim10^2~m_{12}$
from the COM of the binary. On one hand, this offset is much larger than the
size of the binary and hence the wave front is not any more subject to
distortion by the binary--the wave is fully developed \citep{kocsis07}. On the
other, $\Delta r/(2M_3)\lesssim50(m_{12}/M_3)\lesssim10^{-3}$, which means that
it is a small offset relative to the length scale ($2M_3$) of the coordinates.
As a result,  from $r$ to $r+\Delta r$ we do not need to consider the effect of
redshift on the GW amplitude.  Furthermore, the corresponding proper distance,
$\Delta d\simeq \Delta r/\sqrt{1-2M_3/r}<\sqrt{2}\Delta r$, is only a small
fraction of $\lesssim4\times10^{-4}$ of the curvature radius $\rho$, and hence
the intermediate observer is conducting an observation in an effectively flat
space-time. The small redshift and the flatness of space-time allow us to use
the conventional formula $h\simeq4({\cal M}/\Delta d)(\pi f{\cal M})^{2/3}$ to
derive the GW amplitude at the location of the intermediate observer. 

As the wave passes the intermediate observer and propagates towards the final
observer, the amplitude decreases linearly with the proper distance, $d_{p}$.
In the end the final observer detects an amplitude of $ h_o\simeq h(\Delta
d/d_{p})=4({\cal M}/d_{p})(\pi f{\cal M})^{2/3}$, where we have assumed $\Delta
d\ll d_p$. In principle the calculation of $d_{p}$ can be separated into two
parts.  The first part deals with the proper distance close to the SMBH which
is different from the coordinate distance because of the Schwarzschild metric.
The second part is the cosmological comoving distance from the SMBH to the
final observer $d_C$. Since the BBH normally is far from the event horizon of
the SMBH, the cosmological distance is the dominant component. Therefore, we
can neglect the first part of $d_p$ and write $d_{p}\simeq d_C$. Finally, we
find that $h_o$ reduces to that in Equation~(\ref{eq:h_cos}).
 
To complete our analysis, we use the $h_o$, ${\cal M}_o$, and $f_o=f/(1+z_{\rm
tot})$ derived in this section to rewrite Equation~(\ref{eq:d_o}). We find that
\begin{align} d_o&=d_C(1+z_{\rm cos})(1+z_{\rm dop})(1+z_{\rm gra})\\
&=d_L(1+z_{\rm dop})(1+z_{\rm gra}).\label{eq:d_gra} \end{align}
Therefore, the apparent distance $d_o$ is even bigger when the gravitational
redshift is added into the analysis.

\section{The maximum effect}\label{sec:total}

To estimate the upper limit of $z_{\rm tot}$, we need to find the smallest $r$.
For a circular orbit, which would be relevant for the BBHs trapped in AGN
accretion discs, the smallest $r$ is imposed by the innermost stable circular
orbit (ISCO), i.e.  $r=6M_3$ for a non-rotating SMBH.  In this case, the
gravitational redshift is $1+z_{\rm gra}\simeq1.22$ according to
Equation~(\ref{eq:z_gra}).  The circular velocity according to the Keplerian
approximation is $v\simeq c/\sqrt{6}\simeq0.408c$, so the Doppler redshift is
$1+z_{\rm dop}\simeq1.54$ according to Equation~(\ref{eq:z_dop}).  The total
redshift due to these two effects is $(1+z_{\rm dop})(1+z_{\rm
gra})\simeq1.89$,

On the other hand, if the orbit around the SMBH is nearly parabolic, a more
likely configuration in the model of tidal capture, it is the innermost bound
orbit (IBO), i.e. $r=4M_3$, that is limiting $r$. In this case, the
gravitational redshift increases to $1+z_{\rm gra}\simeq1.41$ and a maximum
velocity of $c/\sqrt{2}\simeq0.707c$ can be reached at the pericentre passage,
which corresponds to a Doppler redshift of $1+z_{\rm dop}\simeq2.41$.
Consequently, the total redshift increases to $1+z_{\rm tot}\simeq3.41$. 
 
\begin{figure}
\begin{center}
\includegraphics[width=0.48\textwidth]{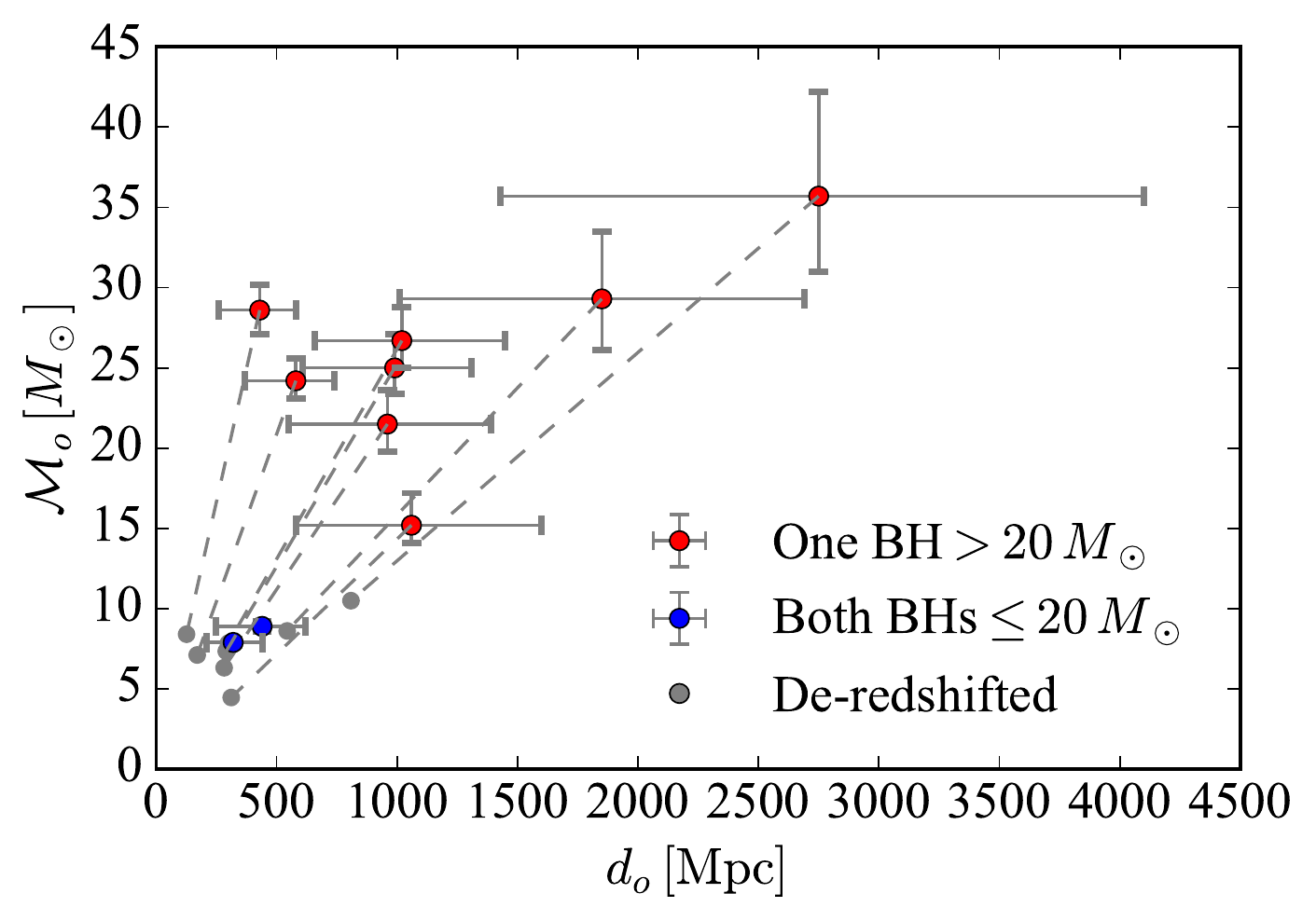} \caption{The observed chirp
masses versus the apparent distances for the ten BBHs detected by LIGO/Virgo
\citep{ligo18}. The red dots refer to the binaries which contain at least one BH
member with a mass greater than $20M_\odot$. The blue ones refer to
those binaries whose BH members are all lighter than $20M_\odot$. The grey dots
attached by dashed lines show the hypothetical chirp masses and distances
derived using the maximum redshift factor of $1+z=3.4$.}
\label{fig:mc} \end{center} \end{figure}

Therefore, the effect caused by Doppler and gravitational redshifts is at most
$1+z_{\rm tot}\simeq(1.9-3.4)$.  Interestingly, these values coincide with the
contrast between the typical mass of the BHs detected by LIGO/Virgo
($20-30M_\odot$) and the canonical mass of the BHs in X-ray binaries
($10M_\odot$). This coincidence can be more clearly seen in
Figure~\ref{fig:mc}, where we plot the observed chirp masses of the ten
detected BBHs against their apparent distances. We color code the data points
according to the masses of the BHs prior to the merger. If one BH member prior
to the merger is greater than $20M_\odot$, we plot the binary as a red dot.
Otherwise, if both BHs are lighter than $20M_\odot$, we plot the binary as a
blue dot.  We choose $20M_\odot$ as the threshold because so far there is no
stringent detection in X-ray binaries of BHs more massive than this mass
\citep{mcclintock14,corral16}. 

\begin{figure}
\begin{center}
\includegraphics[width=0.48\textwidth]{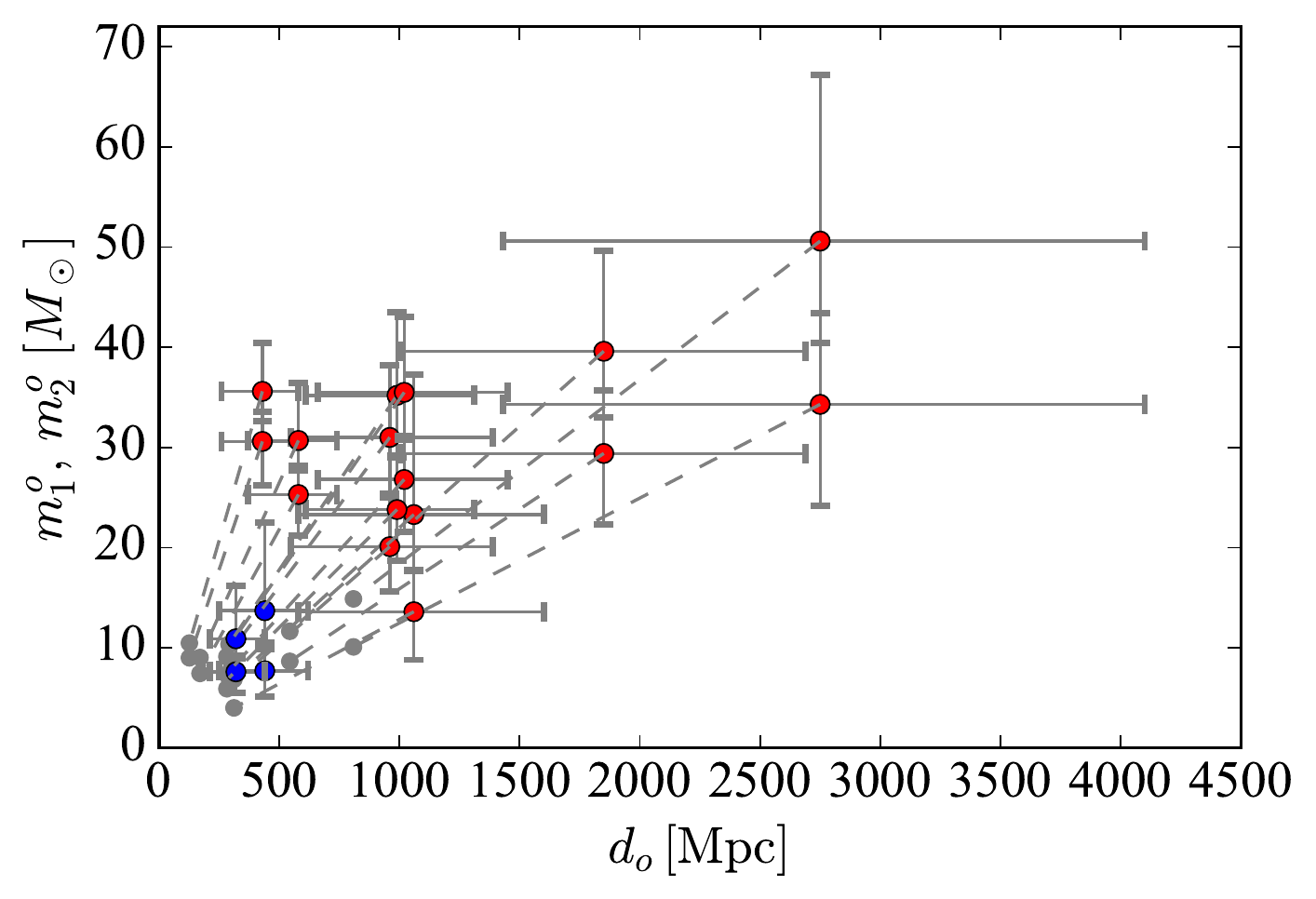} \caption{The same as Figure~\ref{fig:mc} but showing
the mass of each BH prior to the merger.}
\label{fig:m12} \end{center} \end{figure}

The $d_o-{\cal M}_o$ diagram reveals two striking features. (1) There is a gap
between the red and blue populations. It is even more evident when we plot in
Figure~\ref{fig:m12} the mass of each BH prior to the merger. In the light of
our redshift model, a gap is expected because the red population are highly
redshifted, so that they should have been displaced in the diagonal direction,
from the location occupied by the blue population to a region of higher mass
and larger distance. (2) If we use a hypothetical redshift of $1+z_{\rm
tot}=3.4$ to estimate the ``intrinsic'' chirp masses and luminosity distances
for the BBHs in the red population, we find the grey dots in
Figures~\ref{fig:mc} and \ref{fig:m12}.  This grey population occupy the same
region in the diagram as the blue one.  This result is, again, consistent with
the hypothesis that the red population have been redshifted. We note that the
robustness of the features shown here is limited by the current small-number
statistics, but future detection of more BBHs can help us further test our
hypothesis.

\section{Discussion}\label{sec:dis}

In this Letter we have shown that the current astrophysical models allow BBHs
to form and merge at a distance of $r\lesssim10\,R_S$ from a SMBH
(Section~\ref{sec:location}).  In this case, the emitted GWs would be affected
by the Doppler and gravitational redshift in such a away that the mass and
distance derived from the observed waveform are greater than their intrinsic
values by a redshift factor of $1+z_{\rm tot}$ (Sections~\ref{sec:mass} and
\ref{sec:distance}). In the most extreme scenario considered for non-spinning
Schwarzschild SMBHs, $1+z_{\rm tot}$ could be as large as $1.9-3.4$ and,
interestingly, this value agrees with the data of the ten BBHs detected so far
by LIGO/Virgo (Section~\ref{sec:total}).

In the case of non-spinning SMBHs, because the upper limit of $(1+z_{\rm
dop})(1+z_{\rm gra})$ is $3.4$ and the maximum BH mass detected in X-ray
binaries so far is about $20M_\odot$, we expect that the BHs detected by
LIGO/Virgo at relative small distance (so that we know $z_{\rm cos}$ is small)
would not be more massive than $70M_\odot$ before their mergers.  If most SMBHs
are highly spinning, the maximum value of $(1+z_{\rm dop})(1+z_{\rm gra})$
could be even higher, although the exact value deserves a more robust
calculation.  These predictions can be tested against future GW observations. 

Apart from the redshift effect, other types of distortion of the inspiraling
waveform caused by the presence of a SMBH is relatively weak.  (i) The tidal
force of the SMBH is unimportant when the BBH enters the LIGO/Virgo band. This
is so because the duration of the merger event in the band, i.e.  $\Delta t$
derived in the previous section, is much shorter than the Lidov-Kozai timescale
$t_{\rm KL}$ \citep[see][for details]{naoz13}. In fact, in our problem where
$r\sim6M_3$, $a\lesssim10m_{12}$ and $M_3\gtrsim10^6~M_\odot$, $t_{\rm KL}$ is
at least $10^{-2}(M_3/m_{12})^2\simeq3\times10^{7}$ times longer than $\Delta
t$. (ii) The motion of the BBHs around the SMBHs is also insignificant during
the merger.  In principle, the motion induces a time-dependent gravitational
background which leads to a further phase drift of the waveform
\citep{meiron17,inayoshi17}.  But in our case, the merger time $\Delta t$ is
very short, such that during this interval the COM of the binary shifts by only
a length of $v\Delta t\lesssim230m_{12}$, which is more than $400$ times
smaller than the length scale $2M_3$ of the local metric. Therefore, the phase
drift is negligible in our problem.  We note that phase drift becomes important
only when $M_3\lesssim10^4~M_\odot$, because the ratio $v\Delta t/(2M_3)$
becomes greater than $0.2$. But for our problem of $M_3\gtrsim10^6~M_\odot$,
the background is essentially adiabatic. 

The most critical problem of the tidal-capture model is that the event rate
seems too low to explain all the eight massive stellar BHs detected by
LIGO/Virgo. For example, the merger rate is estimated to be at most $0.03\,{\rm
Gpc^{-3}\,yr^{-1}}$ \citep{chen18}, while the rate inferred from the LIGO/Virgo
detections is $(10-10^2)\,{\rm Gpc^{-3}\,yr^{-1}}$ \citep{ligo18}. One possible
mitigation of the tension is that the capture rate of BBHs by SMBH could have
been significantly underestimated.  It is possible because the previous
estimation is based on the (loss-cone) theory which regards BBHs as point
masses. In this way, the diffusion of the binaries in the phase space of energy
and angular momentum around the SMBHs can be studied using the conventional
models developed for single stars.  In reality, however, binaries exchange
energy and angular momentum with background stars in a way more efficiently
than the conventional two-body interactions \citep{heggie75}. For this reason,
the capture rate deserves a re-evaluation.  

The event rate of the BBH mergers in AGN accretion discs is more uncertain.
Possible values range from ${\cal O}(1)\,{\rm Gpc^{-3}\,yr^{-1}}$
\citep{bartos2017,stone2017} to as large as $10^4\,{\rm Gpc^{-3}\,yr^{-1}}$
\citep{mckernan18}. However, the models which lead to these values have not
included the possibility that BBHs could be trapped at a distance as small as
$r\lesssim10R_S$ from a SMBH \citep[][ and also see our
Eq.~\ref{eq:r}]{chakrabarti93}.  We plan to look into this possibility and
calculate the corresponding merger rate in a future work.

\section*{Acknowledgements}

We thank Fukun Liu, Runqiu Liu, Kejia Lee, Rainer Spurzem, Zoltan Haiman, Bence
Kosis and Alberto Sesana for useful discussions.  This work is supported by the
``985 Project'' of Peking University and the NSFC grants No.~11873022,
11773059, 11303039, 11622546 and 11375260.  XC and SL are partly supported by
the Strategic Priority Research Program of the Chinese Academy of Sciences
through the grants No.  XDB23040100 and XDB23010200, and by the Silk Road
Project of the National Astronomical Observatories, Chinese Academy of
Sciences.

\bibliographystyle{mnras.bst}

\bsp    
\label{lastpage}
\end{document}